\documentclass[aps,prl,twocolumn,nofootinbib,showpacs,superscriptaddress,groupedaddress]{revtex4-1}  

\usepackage{dcolumn}
\usepackage{epsfig}
\usepackage{graphicx}
\usepackage{amsfonts}
\usepackage{amssymb}
\usepackage{bm}
\usepackage{latexsym}
\usepackage{amsmath}
\usepackage{hyperref}
\usepackage{color}

\newcommand{\beq}{\begin{equation}}
\newcommand{\eeq}{\end{equation}}
\newcommand{\bea}{\begin{eqnarray}}
\newcommand{\eea}{\end{eqnarray}}

\begin{document}
\setlength{\tabcolsep}{6pt}
\renewcommand{\arraystretch}{1.2}


\title{White Dwarf Critical Tests for Modified Gravity}

\author{Rajeev Kumar Jain}
\author{Chris Kouvaris}
\author{Niklas Gr\o nlund Nielsen}
\affiliation{CP$^3$-Origins, Centre for Cosmology and Particle Physics Phenomenology
University of Southern Denmark, Campusvej 55, 5230 Odense M, Denmark
}

\date{\today}


\begin{abstract}
Scalar-tensor theories of gravity can lead to modifications of the gravitational force inside astrophysical objects. We exhibit that compact stars such as white dwarfs provide a unique set-up to test beyond Horndeski theories of ${\rm G}^3$ type. 
We obtain stringent and independent constraints on the parameter $\Upsilon$ characterizing the deviations from Newtonian gravity 
using the mass-radius relation, the Chandrasekhar mass limit and the maximal rotational frequency of white dwarfs. We find that white dwarfs impose stronger constraints on $\Upsilon$ than red and brown dwarfs. 
\end{abstract}


\maketitle


\section{Introduction}

Scalar-tensor theories of gravity have recently caught a lot of attention as they can serve as a promising candidate for explaining the present cosmic acceleration without the need of a cosmological constant \cite{Clifton:2011jh}.  These theories typically exhibit novel cosmological signatures on very large scales \cite{Jain:2010ka} while being consistent with the predictions of General Relativity on solar system scales, thanks to the Vainshtein mechanism which suppresses the effects of the fifth force on small scales \cite{Vainshtein:1972sx}. {Other analogous screening mechanisms include Chameleon \cite{Khoury:2003aq} and K-mouflage \cite{Babichev:2009ee} which could be effective due to the non-linear structure of the theory}.  Among the large class of these theories, Galileons are well studied examples of the most general theory of a scalar field possessing shift symmetry \cite{Nicolis:2008in}, which ensures that the equations of motion (EoM) are at most second order. As a result, the theory is ghost free. A covariant formulation of Galileons is further required to define them on an arbitrary background in order, for instance, to study their cosmological implications \cite{Deffayet:2009wt}. Covariant Galileons form a subset of the Horndeski theory \cite{Horndeski:1974wa} which is considered the most general theory with a scalar degree of freedom coupled to gravity admitting second order EoM. Recently, it was shown that there also exist healthy extensions of such theories viz. {\it beyond Horndeski theories} \cite{Zumalacarregui:2013pma, Gleyzes:2014dya, Gleyzes:2014qga} which admit self-accelerating solutions thereby being viable competitors to the $\Lambda$CDM model \cite{Kase:2014yya, Barreira:2014jha}. In these theories, the screening mechanism is not completely efficient inside astrophysical objects, in particular, in stars and other compact objects \cite{Kobayashi:2014ida}. Distinct stellar observations have therefore proven to be extremely useful in constraining these theories. 

Within the general class of beyond Horndeski theories of ${\rm G}^3$ type, it has been shown that the Vainshtein mechanism is only partially effective inside astrophysical objects  \cite{Saito:2015fza, Koyama:2015oma}. This leads to a modified equation for the hydrostatic equilibrium
 \begin{equation}
\frac{dP}{dr} = -\frac{Gm\rho}{r^2} - \frac{\Upsilon}{4}G\rho \frac{d^2m}{dr^2}\,,
\label{Eq:Hydrostatic Pressure Original}
\end{equation}
where $P$ and $\rho$ are the pressure and energy density at distance $r$ from the center of the star, respectively and $m$ is the mass enclosed within the radius $r$. The dimensionless parameter $\Upsilon$ characterizes the effects of modifications of gravity and can take arbitrary values. For $\Upsilon>0$, the extra term effectively weakens the gravitational pull inside the star and vice versa. The parameter $\Upsilon$ is further related to the parameters which appear in the effective field theory (EFT) of dark energy as \cite{Gleyzes:2014qga}
 \begin{equation}
\Upsilon = \frac{4 \alpha_H^2}{\alpha_H-\alpha_T-\alpha_B (1+\alpha_T)}\,.
\end{equation}
These different parameters $\alpha_i$ appearing in the EFT of dark energy completely describe the cosmology of the beyond Horndeski theory on linear scales. Note that stable stellar configurations can only exist for $-2/3 <\Upsilon<\infty$ \cite{Saito:2015fza}.  

Recently, it was pointed out that low mass stellar objects such as red and brown dwarfs indeed provide excellent probes of such modifications of gravity \cite{Sakstein:2015zoa}. By utilizing observations of the minimum mass for hydrogen burning in these stars, a strong upper bound was obtained, given by $\Upsilon \leq 0.027$ \cite{Sakstein:2015zoa, Sakstein:2015aac}. It is interesting to note that, the covariant quartic Galileon model \cite{Kase:2014yya} which admits a stable self-accelerating solution for the background expansion leads to $\Upsilon=1/3$ and is therefore ruled out by this constraint. However, red and brown dwarfs cannot provide any constraints on negative $\Upsilon$.
In this letter, we utilize independent observations of compact stars viz. white dwarfs to further constrain the negative regime of $\Upsilon$. White dwarfs prove to be ideal objects for this purpose since their maximum stable mass (the Chandrasekhar mass) is very well known from observations of type 1a supernovae.  We obtain new independent constraints on $\Upsilon$ arising from the observations of the mass-radius relation, the Chandrasekhar mass limit and the maximal rotational frequency of white dwarfs. We find that the mass-radius relation places the most stringent constraint on $\Upsilon$ which for the heaviest observed white dwarf translates to $-0.18 \leq\Upsilon \leq 0.27$. This lower limit on $\Upsilon$ is the strongest so far. 


\section{White dwarf hydrostatic equilibrium}
In a white dwarf, hydrostatic equilibrium is achieved because electrons become degenerate, and the resulting Fermi pressure prevents the star from collapsing under its own gravity. In this section we model a white dwarf in the non-relativistic limit, assuming zero temperature and a chemical composition of fully ionized carbon $^{12}C\,$\footnote{The most abundant carbon and oxygen isotopes $^{12}C$ and $^{16}O$ both have 1/2 electron per nucleon.}.

First, we introduce an equation of state (EoS) to relate pressure and energy density in a white dwarf. The number density of degenerate electrons is $n_e = m_e^3x^3/(3\pi^2)$,
where $x = p_\text{F}/m_e$ with $p_\text{F}$ being the Fermi-momentum and $m_e$ is the electron mass.
The energy density and pressure of electrons are correspondingly given by
$\rho_e = m_e^4 \xi(x)$ and $P_e =m_e^4\psi(x)$ where $\xi$ and $\psi$ are defined in \cite{Shapiro}.
The total energy density is the sum of the energy density of electrons and non-relativistic carbon atoms, $\rho = \rho_e + \rho_C$. However, $\rho$ is completely dominated by $\rho_C$ which is simply related to the number density of electrons as
$\rho_C = m_C n_e/6 = m_C m_e^3 x^3/(18\pi^2)$,
where $m_C$ is the mass of ionized carbon and 6 is carbon's atomic number. Since the pressure from carbon $P_C \ll P_e$, the total pressure becomes $P \approx P_e$. Now we have expressed $\rho$ and $P$  as functions of $x$ which will provide our EoS.
Using  mass continuity $dm/dr = 4\pi r^2 \rho$
we can write Eq.~\ref{Eq:Hydrostatic Pressure Original} as
\begin{equation}
\frac{dP}{dr} = -\frac{Gm\rho}{r^2}\left[ 1+ \frac{\Upsilon  \pi r^3}{m} \left(2\rho + r\frac{d\rho}{dr}\right) \right].
\label{Eq:Hydrostatic Pressure}
\end{equation}
By inserting the expressions for the pressure and energy density, the mass continuity equation and Eq.~\ref{Eq:Hydrostatic Pressure} form a system of two first order ordinary differential equations with unknown functions $m(r)$ and $x(r)$. We integrate this system numerically by using a Runge-Kutta procedure and take the initial conditions to be $m(0)=0$ and $x(0)=x_0$. Here, $x_0$ is related to the central density and pressure through the EoS. The radius $R$ of the star is determined at the point, where the pressure vanishes $P[x(R)]=0$ or equivalently when $x(R)=0$. This also defines the total mass of the star which is $M=m(R)$.


\section{Observational Constraints}

In the following sections we will use the model described in the previous section to constrain the parameter $\Upsilon$ using three independent types of observations of white dwarfs. In our first approach we will compare the mass-radius relation of our model to observations of white dwarfs in binary systems. For our second constraint we use the fact that the mass of a white dwarf cannot exceed the Chandrasekhar limit. Last, we consider rotating stars and require that the centrifugal force does not destroy the stability of the star.


\subsection{Mass-Radius Relation}
The mass and radius of a white dwarf in  binary systems are determined observationally quite accurately. In this section we will compare our model with a  catalogue of twelve white dwarfs compiled in~\cite{Holberg:2012pu}. For these stars both masses and radii and their respective errors are known. Our comparison will consist of a $\chi^2$ test with $\Upsilon$ as a fitting parameter.
\begin{figure}[t!]
\includegraphics[width = .45\textwidth]{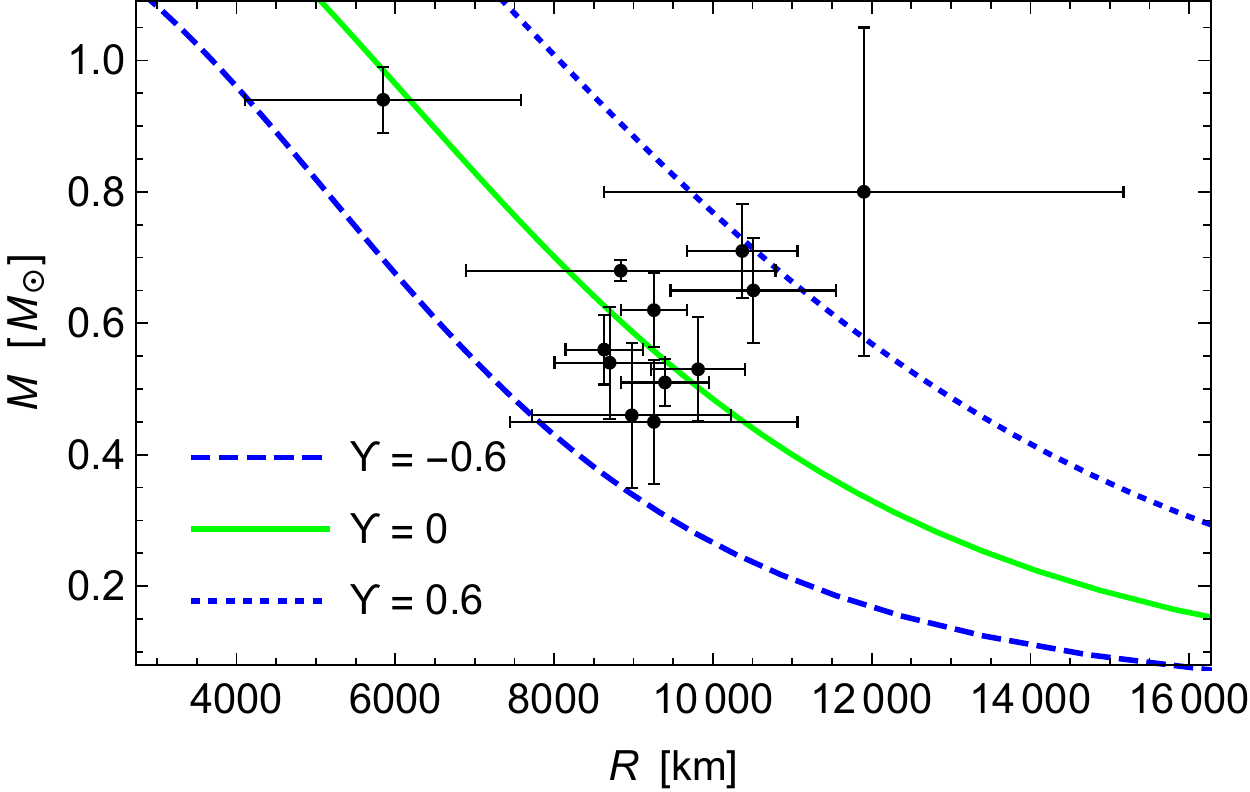}
\caption{White dwarfs from \cite{Holberg:2012pu} with associated error bars. The super-imposed lines are from our theoretical white dwarf model shown for different values of $\Upsilon$. }
\label{Fig:Mass Radius Figure}
\end{figure}
To perform the $\chi^2$ analysis we first determine which point on our theoretical curve $M_\text{th}(R)$ each experimental observation should be compared to. We choose the point on the theoretical curve, which agrees best with each single data point; i.e. we minimize the following quantity
\begin{equation}
\Delta \chi^2_i(R) = \frac{\left(M_\text{th}(R)-M_i\right)^2}{\sigma_{M,i}^2} + \frac{\left(R-R_i\right)^2}{\sigma_{R,i}^2},
\end{equation}
where $M_i, \sigma_{M,i}, R_i$ and $\sigma_{R,i}$ are the mass, mass standard deviation, radius and radius standard deviation, respectively, of the $i\,$th star. The final $\chi^2$ is thus
\begin{equation}
\chi^2 = \sum_{i=1}^N \Delta\chi_i^2(\mathcal{R}_i),
\end{equation}
where $\mathcal{R}_i$ is the value of $R$ that minimizes the corresponding $\Delta\chi_i^2$.
For a good fit  $\chi^2/\text{d.o.f.}$ should be less than one. The d.o.f. is the number of degrees of freedom, which in our case is $\text{d.o.f.} = 2N - n - 1 = 22$. Here $N=12$ is the number of stars, with the factor of 2 coming  from the fact that we have two independent observations for each star (i.e. radius and mass) and $n=1$ is the number of fitting parameters in our case.

\begin{figure}[t!]
\includegraphics[width = .45\textwidth]{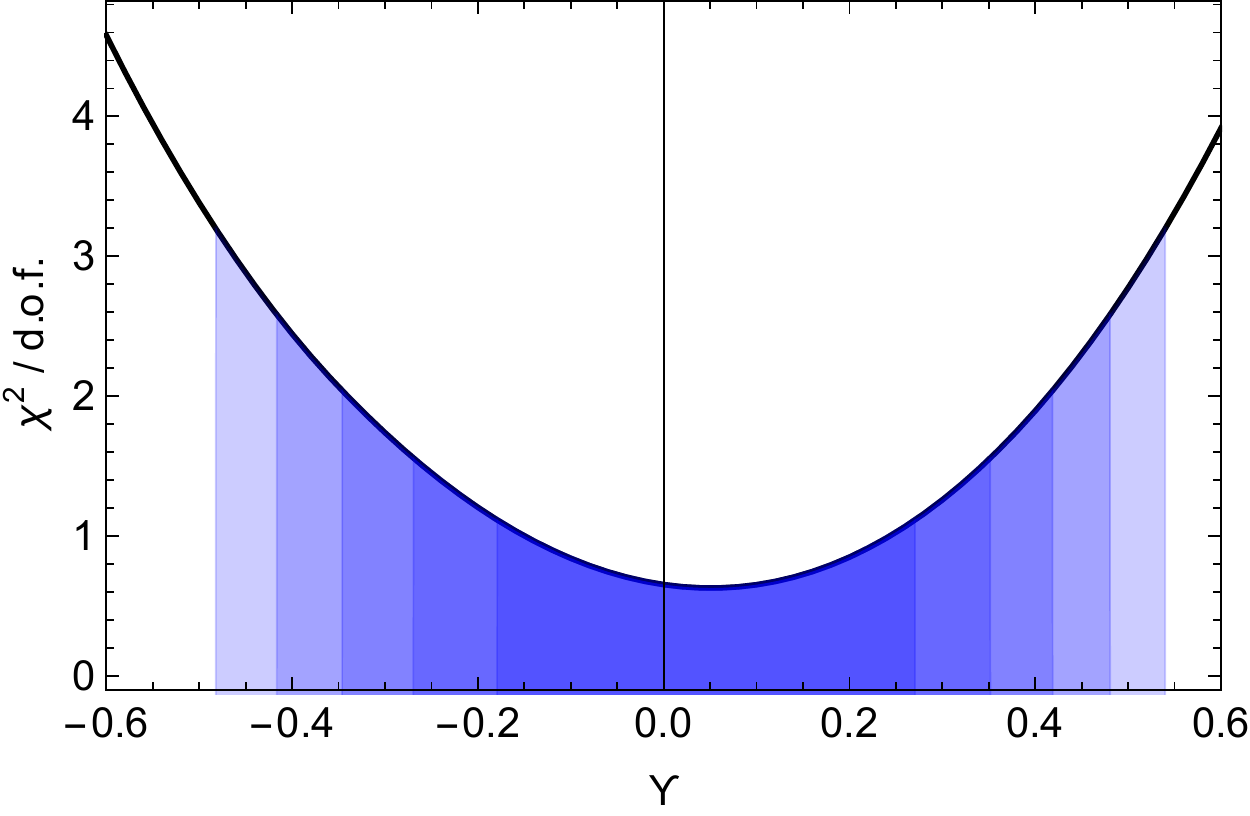}
\caption{The curve is $\chi^2/\text{d.o.f.}$ with $\text{d.o.f.}=22$. The shaded blue regions are 1 to 5$\sigma$ confidence levels. The darkest shaded region is $1\sigma$ and the lightest is $5\sigma$.}
\label{Fig:chi2}
\end{figure}

In Fig.~\ref{Fig:Mass Radius Figure} we  show the masses and radii of white dwarfs from the catalogue in~\cite{Holberg:2012pu}. We have super-imposed $M_\text{th}(R)$ for our model with different values of $\Upsilon$. In Fig.~\ref{Fig:chi2}  we plot $\chi^2/\text{d.o.f.}$ as a function of $\Upsilon$ and also show confidence levels for the consistency of modified theories with respect to observations for the parameter $\Upsilon$ up to $5 \sigma$.
From Fig.~\ref{Fig:chi2} we see that the $\chi^2/\text{d.o.f.}$ is minimized around $\Upsilon=0$, i.e. our model is fully consistent with the data in the absence of modified gravity. Switching on a positive or negative $\Upsilon$ starts creating a  tension with the data to the point where {the allowed range for $\Upsilon$ falls within $-0.18 \leq \Upsilon \leq 0.27$ at $1\sigma$ and $-0.48 \leq \Upsilon \leq 0.54$ at $5\sigma$.} Since we use a fairly small catalogue of stars, these limits can conceivably be made  stronger in the future. 
The analysis in this section can also be improved by taking into account the specific temperature of each observed star, when the theoretical mass-radius relation is computed. Of the 12 stars we consider, 3 stars have temperatures higher than 20,000 K \cite{Holberg:2012pu}. Above this temperature the radius may be significantly affected \cite{Barstow15}.


\subsection{Chandrasekhar Mass Limit}
Here we  examine the effect of a non-zero $\Upsilon$ on the Chandrasekhar limit of a white dwarf. Since $\Upsilon>0$ will effectively make gravity weaker inside astrophysical objects, the Chandrasekhar mass limit will increase, as the degeneracy pressure can support more matter. Conversely, if $\Upsilon$ is negative gravity will effectively be stronger and the Chandrasekhar limit will decrease.

\begin{figure}[t!]
\includegraphics[width = .45\textwidth]{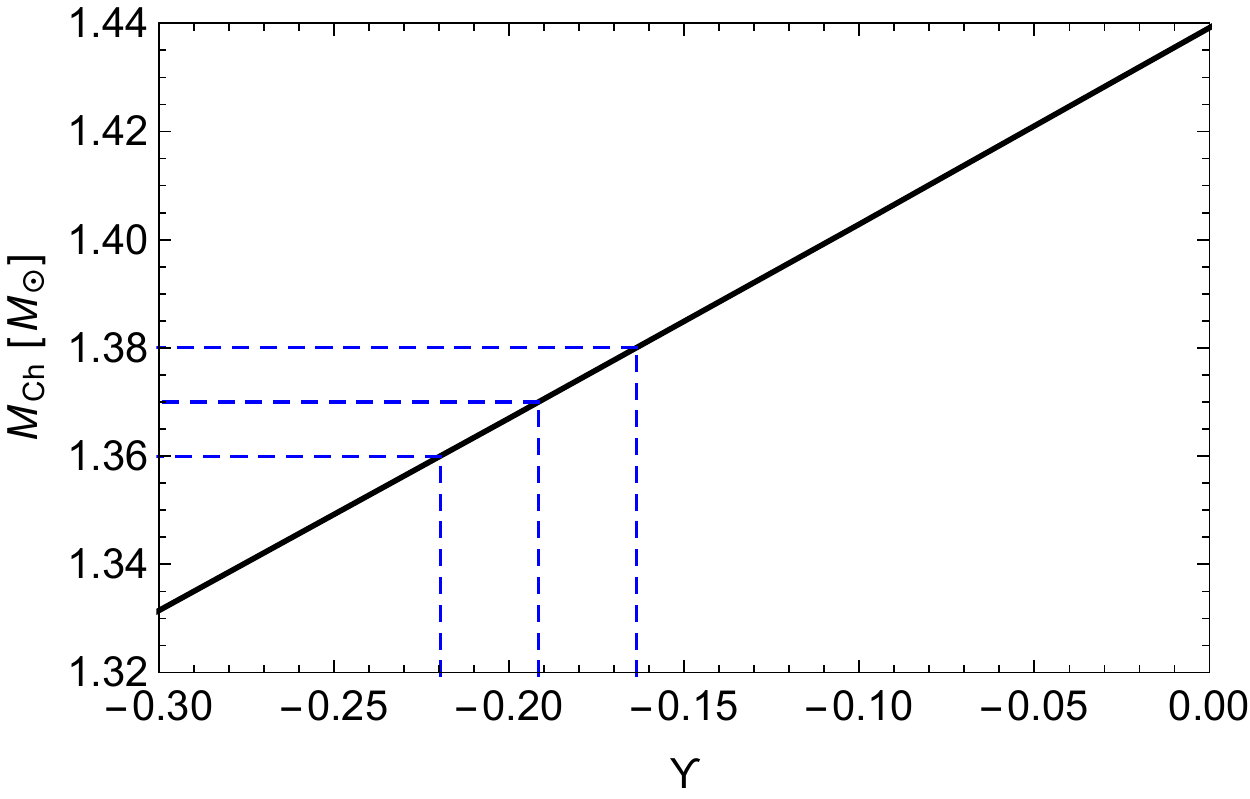}
\caption{The Chandrasekhar mass (solid black line) is plotted as a function of $\Upsilon$. The blue dashed lines indicate the lower limit on $\Upsilon$ which corresponds to the heaviest estimated white dwarf of $1.37\pm 0.01M_\odot$ \cite{Hachisu:2000nx}.}
\label{Fig:Chandrasekhar Figure}
\end{figure}

The limit obtained in this section will be akin to that of \cite{Sakstein:2015zoa, Sakstein:2015aac}, wherein the minimum mass for burning hydrogen in a red dwarf was used to place a strong upper limit on $\Upsilon$. Our approach differs since we use white dwarf stars and place the most stringent limit on negative $\Upsilon$ instead.
The Chandrasekhar limit in our model is $1.44 M_\odot$. This value is a few percent larger than that of numerical calculations that take into account corrections from general relativity, a detailed chemical composition of the star, non-zero temperature etc. \cite{Nomoto:1982zz, Pacini:1987ec}. Requiring the heaviest known white dwarf to be lighter than the Chandrasekhar mass sets a lower (negative) limit on $\Upsilon$ in our model as illustrated in Fig.~\ref{Fig:Chandrasekhar Figure}. Many white dwarfs have been observed with masses $>1.3M_\odot$ \cite{Nalezyty04}. To the best of our knowledge the heaviest estimated white dwarf in the literature appeared in \cite{Hachisu:2000nx} with a mass of $(1.37\pm 0.01) M_\odot$. The 1$\sigma$ limit on $\Upsilon$ that we obtain from this white dwarf is $\Upsilon \geq-0.22$. 

We can further limit $\Upsilon$ from above by assuming that the Chandrasekhar mass of standard non-rotating white dwarfs cannot be arbitrarily large. We will however not place such a theoretical limit here, since we already have obtained upper limits of comparable strength from observations of the mass-radius relation.

The analysis in this section differs from that of the mass-radius relation by being independent of radius measurements. For the Chandrasekhar limit the heaviest star sets the exclusion limit on $\Upsilon$, whereas the mass-radius relation limit relies on statistics and can be improved by increasing the number of observations or better determination of the errors.
Other modifications of gravity as in $f(R)$ theories have also been shown to affect the Chandrasekhar mass limit \cite{Das:2015gla,Das:2014rca}.


\subsection{Rotational Frequency}
A star rotating with angular frequency $\omega$ receives a positive contribution to the pressure due to the centrifugal force. In this section we will examine the constraints on $\Upsilon$ set by fast spinning white dwarfs. We approximate that a rotating white dwarf remains a sphere, and that the rotation period is constant throughout the star. Within these approximations we need only to append Eq.~\ref{Eq:Hydrostatic Pressure} by a centrifugal term to include the effect of rotation of the star
\begin{equation}
\frac{dP}{dr} =-\frac{Gm\rho}{r^2}\left[ 1+ \frac{\Upsilon  \pi r^3}{m} \left(2\rho + r\frac{d\rho}{dr}\right) \right] +  \rho\, \omega^2 r.
\end{equation}
At any radius within the star, $dP/dr$ must remain negative. For a given central density and $\omega$, the parameter $\Upsilon$ is constrained by this negativity condition. The constraint can be written as
\begin{equation}
\Upsilon > \left(-\frac{4}{3} \frac{\rho_\text{avg}}{\rho}+\frac{1}{\pi}\frac{\omega^2}{G\rho} \right)\left( 2 + \frac{d\log\rho}{d\log r}\right)^{-1},
\label{Eq:Rotation bound 1}
\end{equation}
if $(2+d\log \rho/d\log r) >0$ and where $\rho_\text{avg}(r) \equiv m(r)/(4\pi r^3/3)$. For $(2 + d\log \rho/d\log r )<0$ the inequality changes sign. Note that in the case of constant density {and $\omega \to 0$}, the above constraint reduces to $\Upsilon>-2/3$, which was found as a universal lower bound in \cite{Saito:2015fza}\footnote{In the notation of \cite{Saito:2015fza} it is an upper bound.}. The constraint in Eq.~\ref{Eq:Rotation bound 1} can therefore be considered as a generalization of this bound. When $(2 + d\log \rho/d\log r )<0$ is satisfied the constraint becomes an upper bound. Since $\rho$ is a monotonically decreasing function of $r$, this can be satisfied if $\rho$ is steeper than $r^{-2}$ at some radius inside the star.

Non-zero values of the parameters $\omega$ and $\Upsilon$, will back-react on the right-hand sides of Eq.~\ref{Eq:Rotation bound 1}. Consequently, the bounds depend on the particular star.  As one would expect, the strongest bounds come from the fastest rotating systems. While white dwarfs usually rotate  slowly compared to the maximal possible rotational frequency, there do exist fast rotating white dwarfs. Such an example is \emph{RX J0648.0--4418} \cite{Mereghetti:2011bh} which has a period of 13.2 s and a mass of $(1.28 \pm 0.05) M_\odot$.

\begin{figure}[t!]
\includegraphics[width=0.45\textwidth]{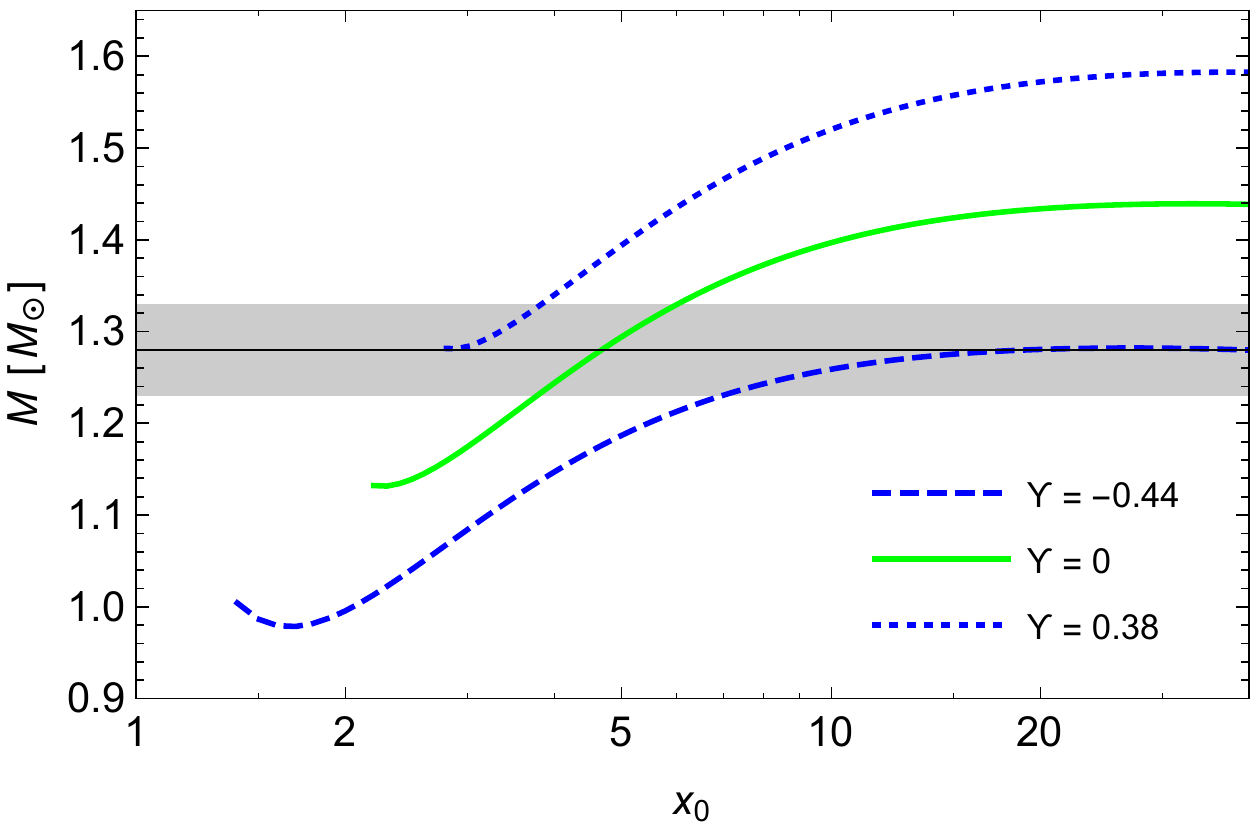}
\caption{The stellar mass as a function of the Fermi momentum $x_0 = p_\text{F}/m_e$ of electrons in the center of the white dwarf. The period of rotation is taken to be 13.2 s, and the black line and gray shaded region are $1.28 \pm 0.05 M_\odot$ corresponding to \emph{RX J0648.0--4418} \cite{Mereghetti:2011bh}.}
\label{Fig:Rotation figure}
\end{figure}

For a given $\Upsilon$, any non-zero $\omega$ will lead to a minimum stellar mass for which hydrostatic equilibrium can be upheld. Below the minimum mass, the centrifugal force is always stronger than the self gravity of the star. Furthermore, as discussed in the previous section the mass of the star can never exceed the Chandrasekhar limit. In the case of a particular observed star, we can therefore ask how large $\Upsilon$ can be, if the observed mass is to be consistent with the observed rotational frequency. This is illustrated for \emph{RX J0648.0--4418} in Fig.~\ref{Fig:Rotation figure}. Here we have scanned the stellar mass over a range of $x_0$ which is related to the central density and pressure by means of EoS. At low $x_0$ the lines terminate, since the resulting mass cannot sustain such fast rotations. We see that the interval $-0.44 \leq \Upsilon \leq 0.38$ can be consistent with a stellar mass of $1.28 M_\odot$, whereas larger or smaller values either produce too massive or too light stars. The observational $1\sigma$ error on the mass of \emph{RX J0648.0--4418} is $\pm 0.05 M_\odot$. Considering this error, the allowed value of $\Upsilon$ falls within $-0.59\leq\Upsilon\leq 0.50$.

We should mention here that this is again a conservative limit. We assumed that the white dwarf rotates with a uniform angular velocity. However, in reality, it has been indicated that the inner parts of a white dwarf rotate faster~\cite{Kawaler:1998fc}. This would make the constraint on $\Upsilon$ even tighter.

\section{Conclusions}

In this letter, we have shown that compact stars such as white dwarfs provide a unique laboratory to probe the small scale imprints of beyond Horndeski theories. We found that among all the observational properties of white dwarfs, the strongest limit on $\Upsilon$ arises from the mass-radius relation {$-0.18 \leq\Upsilon \leq 0.27$}. This stringent constraint further reduces the viable parameter space of these alternative gravity theories. 
The lower bound obtained here improves the previous limit on $\Upsilon$ in the literature by a factor of more than 3. We also obtain independent limits on $\Upsilon$ by considering the Chandrasekhar mass limit and the stability of fast rotating white dwarfs. 

\vskip 6pt\noindent
{{\it Acknowledgments.}~The CP$^3$-Origins center is partially funded by the Danish National Research Foundation, grant number DNRF90.


\end{document}